# Development of a xenon triple point apparatus suitable for calibrating long-stem SPRTs and preliminary measurements of the temperature

Yikun Zhao[a,b,c,d], Jintao Zhang[1b], Xiaojuan Feng[1b], Yu Liang[b], Yongdong He[1a], Hua Zhuo[c,d], Xiangrui Deng[b,c], and Haibing Li[c]
[a]School of Physics Science and Technology, Xinjiang University, Urumqi 830017, PR China
[b]National Institute of Metrology, Beijing 100029, PR China
[c]Xinjiang Uygur Autonomous Region Research Institute of Measurement and Testing, Urumqi 830011, PR China
[d]Xinjiang Key Laboratory of Agricultural Unmanned Aircraft Performance and Safety, Urumqi 830011, PR China

## Abstract

Xenon is of high chemical-physical stability and health compatibility. The xenon triple point (Xe TP) is accounted for a promising candidate replacing the mercury triple point (Hg TP) from the set of the defining fixed points of the international temperature scale ITS-90. The success of the alternative highly depends on the level of the realization of the Xe TP using long-stem standard platinum resistance thermometers (LSPRTs). We report in this article our study on the development of an immersion-type Xe TP apparatus, which is suitable for calibration of both LSPRTs and capsule standard platinum resistance thermometers (CSPRT). We realize the melting plateaus of the Xe TP using the continuous heating method on the apparatus. The effective melting plateaus extend for 8-12 hours long with temperature flatness range of 0.37 mK−1.0 mK over the melted fractions from 0.2 to 0.75. We find the axial heat leak contributing a principal effect influencing measurements of the Xe TP. We investigate the effect by varying the offset temperatures on the outer wall of the Xe TP cell. We measure the Xe TP using two LSPRTs upon correction of the axial heat leak. The new measurement, giving the Xe TP of 161.405 71 (55) K ($k$=1) at the melted fraction $F$=1.0, agrees well with those previously obtained by the adiabatic apparatuses. Their differences fall within 0.11 mK to 0.42 mK. by. Those differences are well covered by the estimated measurement uncertainty.

[1] The corresponding authors: zhangjint@nim.ac.cn, fengxj@nim.ac.cn, hydongq@126.com

## 1. Introduction

In order to reduce the Type 1 inconsistency (SRI) and Type 3 non-uniqueness (NU3) of subrange 4, the mercury triple point (Hg TP) (234.3156 K) was adopted by the International Temperature Scale of 1990 (ITS-90) as the unique defining fixed point between the argon triple point (Ar TP) and water triple point (TPW) (83.8058 K to 273.16 K) [1,2]. Literature estimates that the Hg TP leads to a reduction of ±0.5 mK for the effect of SRI and NU3 [3]. However, because the temperature of the Hg TP is close to that of the TPW, the asymmetry makes the interpolation of the ITS-90 between the Ar TP and the Hg TP highly sensitive to the measurement uncertainty of the Hg TP [4]. Further, mercury is highly toxic and incompatible with environment. The Minamata Convention enforces prohibition of the commercial trade and international transportation of mercury [5]. Therefore, finding a proper candidate replacing the Hg TP is important. The noble gas xenon has excellent chemical-physical stability and low health risks. The xenon triple point (Xe TP) has a nominal temperature of 161 K, being at the midway between the Ar TP and TPW. Those properties make the Xe TP a preference candidate replacing the Hg TP. If replacement is done, the effect of NU3 with standard platinum resistance thermometers (SPRTs) is estimated to be reduced from ±0.5 mK to ±0.1 mK [6].

Michels and Prins [7] studied the melting plateaus of the Xe TP in 1962 and reported the temperature of the Xe TP as 161.385 K [7]. The Xe TP was initially considered as a candidate of the defining fixed points of the ITS-90 [8-17]. Unfortunately, the studies at the time demonstrated a metrology performance of the Xe TP inferior to that of the Hg TP. The Xe TP was finally removed from the candidate list. Researchers attributed the apparent inconsistency of the Xe TP to the effect of impurity and isotope. In 1996, CCT proposed including the Xe TP in list of the first-quality secondary reference fixed-points with a recommended standard uncertainty of 1 mK for a sample purity of 99.995% (4N5) [18]. In 2005, Hill and Steele [19] reported a new measurement of the Xe TP using a 7N purity sample and adiabatic means through a sealed cell suitable for capsule standard platinum resistance thermometers (CSPRTs). They claimed an uncertainty of 0.32 mK, remarking a significant progress compared with the previous record. The small uncertainty is comparable to that of the Hg TP [19, 20]. In 2014, Petre Steur *et al*. [21] repeated the realization using the same batch of sample and the adiabatic procedure. The new realization claimed the uncertainty of 0.27 mK. The results agree well [22]. The new efforts demonstrate the possibility to replace the Hg TP by the Xe TP. Kawamura *et al*. [23] reported the most recent result of 161.40613 K by the adiabatic procedure. The modest difference from literature [19,20] lies in the used samples. The result given in literature [23] differs from literature [19,20] by 0.17-0.31 mK. We are motivated by the new progresses to extend the research suitable for LSPRTs. It is well known that LSPRTs represent the principal interpolation instruments for ITS-90 in the subrange. A thorough study of the Xe TP using LSPRTs has an appreciable significance on the campaign replacing the Hg TP. Thus, we aim at developing an immersion-type Xe TP apparatus suitable for LSPRTs of a comparable uncertainty and consistency with those by the adiabatic means.

We report in this article our design of an immersion apparatus for realizing of the Xe TP. The sample cell is accommodated in a vacuum jacket of two radiation screens. The assembly of cell and jacket is immersed in liquid nitrogen. The cell has the thermometer well guiding LSPRTs from the room environment into the cell at the Xe TP. The apparatus runs in the quasi-adiabatic principle. We obtain the melting plateaus of effective length about 8-12 hours with temperature flatness range of 0.37 mK–1.0 mK between $F$=0.2 and $F$=0.75. We find the axial heat leak playing a major factor influencing measurements of the Xe TP. Upon correction of the axial heat leak, our new results agree well with those attained through the adiabatic apparatus [19,21,23].

## 2. Experimental sample and apparatus

### *2.1. Purity of xenon gas sample and design of immersion cell*

The xenon sample used in this work is produced by Linde Corporation with a claimed purity 5N5. The producer claims the chemical analysis report in Table 1. Reference [19] indicates that impurity Kr is the root cause affecting the Xe TP temperature. The content of Kr is reported of 0.1 parts per million (0.10 ppm).

**Table 1**

Chemical analysis of the xenon sample

| Impurity | Analytical results (ppm) |
|---|---|
| $O_2$ | 0.10 |
| $N_2$ | 0.11 |
| THC | 0.10 |
| $H_2O$ | 0.10 |
| $CO_2$ | 0.10 |
| CO | 0.10 |
| $CF_4$ | 0.10 |
| Ar | 0.10 |
| Kr | 0.10 |
| $H_2$ | 0.10 |
| Total | 1.01 |

In order to deplete the impurities of lower freezing points than that of Xe TP, we pump the cell after the Xe sample has been frozen into solid mantle. According to the partial pressures, the impurities of THC, $H_2O$ and $CO_2$ in the Xe sample will be solidified when exposed to liquid nitrogen. The other impurities are all in their gas phase. The partial pressures of the gaseous impurities, each of less than 0.1 parts per million (0.1 ppm), are estimated in 0.719–0.987 Pa. We pump the cell to a low pressure in $10^{-4}$ Pa. We repeated the melting-frozen-pumping process in twice. We anticipate the major contents of the gaseous impurities being depleted from. The impurities of THC, $H_2O$ and $CO_2$ are the rest of the total content in 0.3 ppm. The inverse first cryogenic constant for xenon is $A^{-1}$=93.6 μK·ppm$^{-1}$. Based on the overall maximum estimation (OME) model [24], the remaining impurities affect the temperature of the triple point by 0.055 mK at $F$=0.5.

Xenon is composed of multiple stable isotopes. Among those, $^{132}Xe$ (26.89%) and $^{129}Xe$ (26.27%) dominate the character of the triple point. The maximum estimated uncertainty arising from xenon's isotope effect is only 21 μK, as reported in the literature [25].

## 2.2. Experimental setup

We aim at developing an immersion-type apparatus appropriate to calibration of LSPRTs. To meet the objective, the sample cell has a deep thermometer well accommodating the sensing element and the associate stem of SPRT. Strous *et al*. [26] and Furukawa *et al*. [27] recommend minimum depth of thermometer well above 155 mm. The depths in 180 mm–220 mm are optimal. In accordance with the recommendation, the cell shall contain a bulk sample at the triple point. The critical temperature of xenon is 289.73 K [28]. Thus, the Xe sample in the cell will be in supercritical state at the room temperature. The sample pressure is extremely sensitive to ambient temperature. A minor increase of ambient temperature causes a large effect in sample pressure. For operation safety and facility, a low pressure is to be preferred. For general cylinder cell, the requirements of sufficiently long thermometer well are incompatible with the requirement of proper pressure at the room temperature [29]. By referring to literature [30,31] about the $CO_2$ TP, we design a cell of the double-cylinder structure with a trapezoidal connection as shown in Fig. 1. Our design is similar to that of the $CO_2$ TP cells, but there are differences. Considering comprehensively the heat leakage of the xenon triple point cell and the long-stem thermometer under low-temperature conditions, the high-pressure safety of the sample at room temperature, and the operability of calibrating the long-stem thermometer, the total design height of the Xe TP cell is slightly higher than that of the $CO_2$ TP cell, including a greater immersion depth and a larger gas buffer chamber. The upper cylinder has a radial extension allowing for a larger expansion volume of gas phase [31]. Under a lower pressure compared with that in a single cylindrical cell, more samples can be accommodated, ensuring that the immersion depth of LSPRT in the cell exceeds the recommended depth of 155 mm [26] and [27].

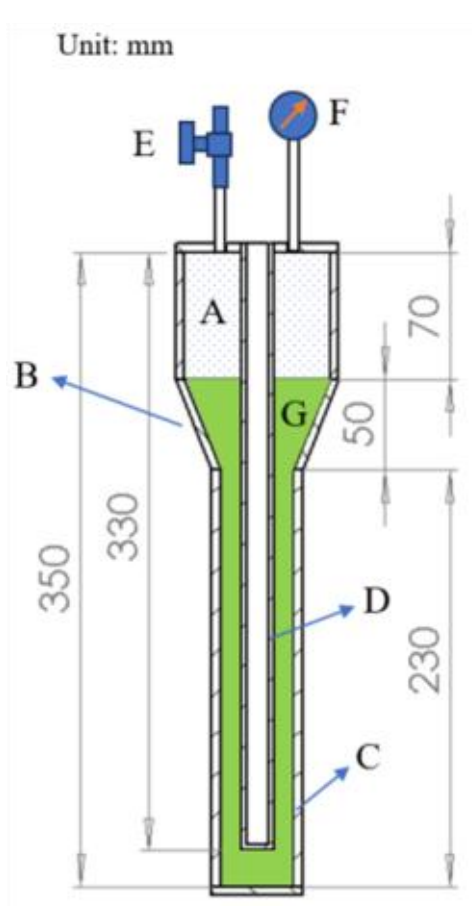

**Fig. 1.** Sketch of the immersion-type Xe TP cell: (A) gas buffer chamber; (B) transition section; (C) small cylindrical cavity ; (D) central thermowell; (E) bellows valve; (F) pressure gauge and (G) condensed portion of xenon (shown in green).

The internal height of the cell is 350 mm. The cell is divided into three parts. The upper cylinder has a height of 70 mm and a diameter of 80 mm. The trapezoidal connection has a height of 50 mm. The lower cylinder has a height of 230 mm and a diameter of 40 mm. The central thermometer well has a height of 330 mm. The design offers a volume of 703.5 ml for containing sample. In Fig. 1, the symbols E, F, and G denote a bellows valve, a pressure gauge, and a solidification section of Xe sample (shown in green), respectively.

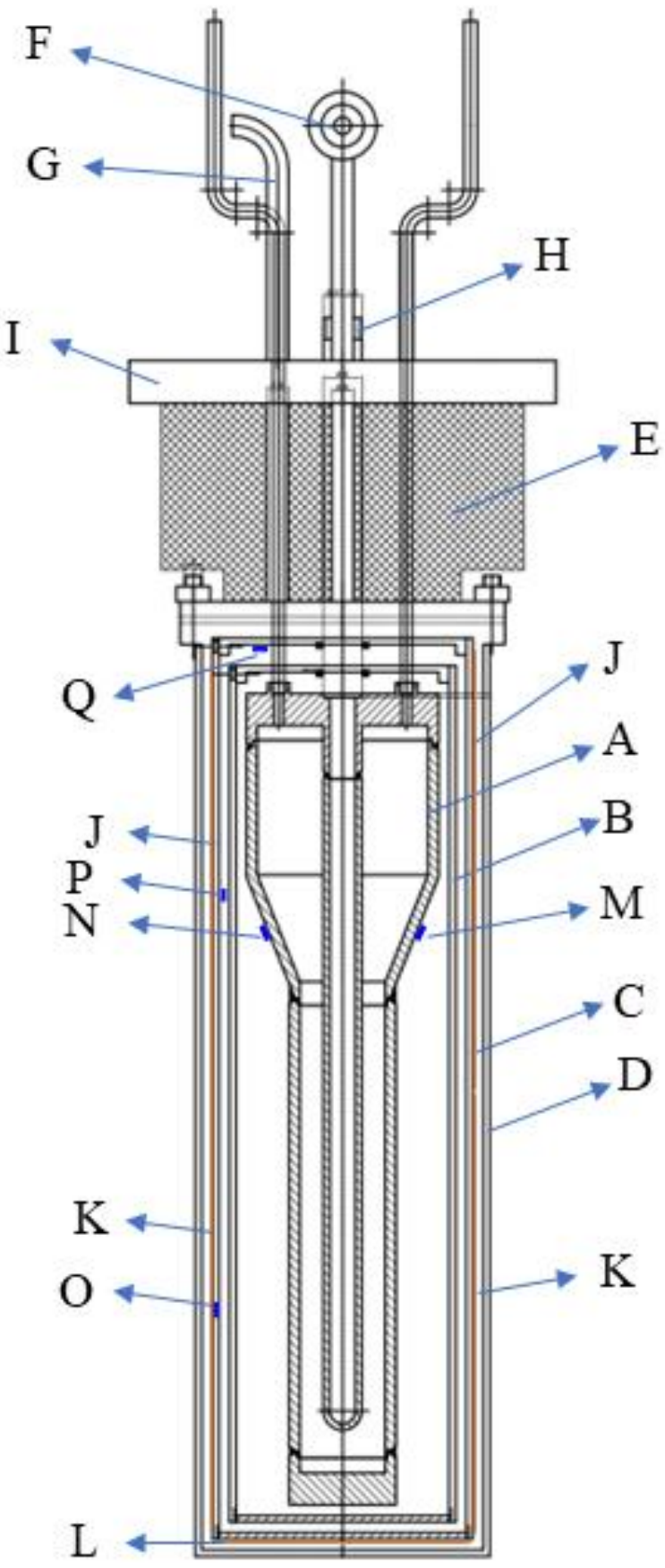

**Fig. 2.** Cross section of the immersion Xe TP apparatus：(A) Cell; (B) internal radiation screen; (C) external radiation screen; (D) vacuum chamber; (E) thermal insulation materials; (F) vacuum chamber vacuum pumping and He filling tube; (G) lead tube; (H) central thermowell; (I) flange; (J) - (K) cylindrical thin film heater; (L) the circular thin film heater and (M) - (Q) numbers are A-01 #, A-02 #, A-03 #, A-04 #, and A-05 # Pt100 thin film platinum resistance sensors, respectively.

The cross section of the immersion Xe TP apparatus is pictured in Fig. 2. The apparatus is composed of 17 elements labelled (A)-(Q). (B) and (C) account for the inner and outer radiation screens, respectively. They are made of oxygen-free high-conductivity (OHFC) copper in cylindrical canes. Their surfaces are plated in gold for reduction of radiation heat. The outer screen is temperature-controlled, (D) labels the vacuum jacket that maintains vacuum during the process of realizing the triple point. When we solidify the sample and initiate melting plateaus, He is filled into the jacket to enhance heat exchange. (E) denotes thermal insulation layer isolating cell from outer ambient. (H) stands for a central thermowell with an inner diameter of 12 mm. The well of such large diameter allows calibrating both LSPRT and CSPRT. (see references [32-34] for calibration of CSPRTs). (J) denotes a film heater applied with the upper half of the external radiation shield. (K) and (L) stand respectively for a circular and an annular heater. They connect in series heating the lower half and endplate of the external radiation shield. (M)-(Q) labels A-01#, A-02#, A-03#, A-04#, and A-05# thin-film platinum resistance thermometers of nominal resistance in 100 Ω. A-3# and A-5# monitor respectively the temperature of the upper and lower sections of the external shield and provide entries for the corresponding controllers; A-1#, A-2#, and A-4# monitor the temperature on the surfaces of the cell's outer wall and the lower surface of the top flange of the external radiation shield, respectively.

*2.3. Gas-handing Manifold*

By referring to [31], we design the gas-handing manifold sketched in Fig. 3. All components of the manifold are made of stainless steel, including stainless steel electropolished tubes, valves and connectors.

The Xe cell is ultrasonically cleaned before being connected to the gas-handling manifold. We first clean the cell in anhydrous ethanol. We then immerse the cell in the mixture of anhydrous ethanol and acetone of a volume ratio in 4:1. After cleaning, we connect the cell to the manifold. We degas the manifold and the cell in one month by persistent pumping and baking using a molecular pump. The baking temperature is 140 °C. During the period, we flush the manifold and the cell using pure Xe gas in twice. Each flushing lasts 12 hours. We fill the manifold and the cell with pure Xe gas at 200 kPa–250 kPa to remove trace gases absorbed on walls of manifold components and cell.

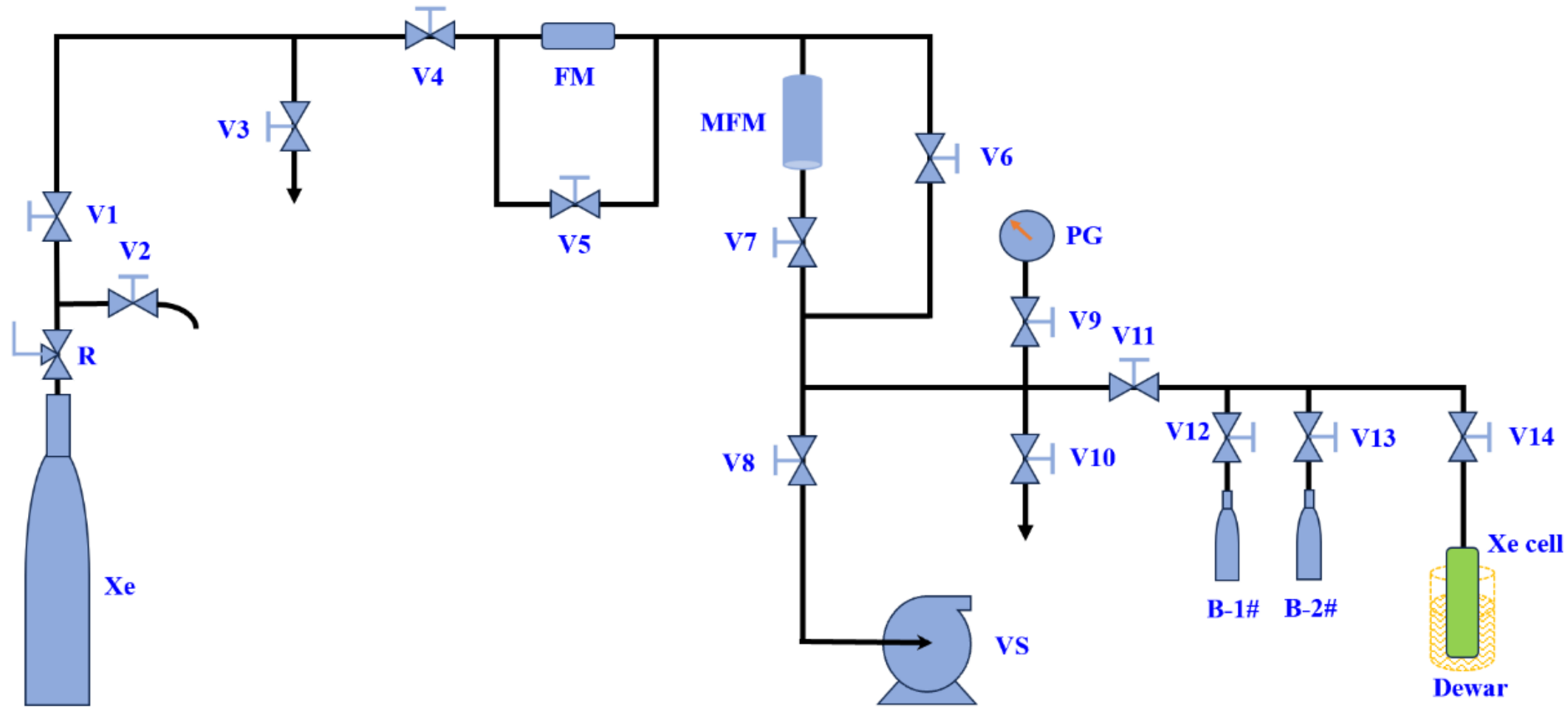

**Fig. 3.** The gas-handling manifold used for filling Xe cell. From left to right: Xe source cylinder, Xe; pressure regulator, R; V1 to V14 are valves; flow meter, FM; mass flow meter MFM; "molecular pump-mechanical pump" two-stage pump vacuum system, VS; pressure gauge, PG; No.1 sampling bottle, B-1#; No.2 sampling bottle, B-2#; Xe triple-point cell, Xe cell; liquid nitrogen dewar, Dewar.

Upon the degassing process, we fill the cell with Xe with an amount of 1141.6 g. The cell has a volume of 707 ml. Thus, the total density of Xe in the cell is 1614.7 kg·m$^{-3}$. In accordance with the density, we tabulate in Table 2 our estimation of the ambient temperature dependence of the sample pressure. Our cell was designed to endure a pressure of 15 MPa. We estimate the sample pressure of 7.1 MPa at 20 °C and 14.0 MPa at 50 °C. We generally store the cell below 30 °C. Thus, the sample pressure is within the safe limit. Besides, the installed pressure gauge on the cell indicates a pressure reading of 7.1 MPa at 20.1 °C. Using the NIST Standard Reference Database Program [28], we calculate our sample density as 1611.1 kg·m$^{3}$, which agrees with that of our weight of the xenon sample in the cell.

In accordance to the filling mass, the amount of 99.7% Xe sample is in solid and liquid phases when the cell is cooled to the Xe TP. The condensed portion of xenon provides an effective immersion depth of 222-238 mm for LSPRTs. Such an immersion depth exceeds at least the minimum value of 155 mm recommended by Strouse [26] and Furukawa [27].

**Table 2**

The temperature dependence of pressure in cell.

| Temperature/°C | 0.0 | 10.0 | 20.0 | 30.0 | 40.0 | 50.0 |
|---|---|---|---|---|---|---|
| Pressure/MPa | 4.1 | 5.1 | 7.1 | 9.4 | 11.7 | 14.0 |

*2.4. Experimental system*

We design the experimental setup for realization of the Xe TP. Its main feature is drawn in Fig. 4. A liquid nitrogen Dewar is used accommodating the vacuum jacket in Fig. 2. The liquid nitrogen provides a cooling environment for the vacuum jacket assembly. Two platinum resistance thermometers are attached on the upper and lower part of the temperature-controllable radiation screen, respectively. The readings of the thermometers are fed to two temperature controllers, respectively. The upper and lower part of screen is independently controlled with a temperature resolution of 0.01 mK. Besides, we monitor the outer wall of the triple point cell using two other platinum resistance thermometers. The two thermometers are symmetrically mounted on both sides of the cell's outer wall. We average their readings as the temperature records of the outer wall of cell.

The Dewar of 94 L is filled with liquid nitrogen by a self-pressurized tank. The Dewar is filled to a depth level of 90–95% before operation for the Xe triple point. During operation, the Dewar is decoupled switching off from the tank. From our experimental experience, we recognize the minimum level to be 30% as stated in Section 4.1. We apply a gauge monitoring liquid level. For the experimental setup, the duration is about 29 hours for liquid nitrogen being consumed from 91% to 32%. During the operation, we protect the thermometer well from icing by filling the well with nitrogen gas. The nitrogen gas is supplied by the nitrogen cylinder shown in Fig. 4. We apply a silicone rubber O-ring sealing the stem of SPRT inside the well. Thus, the nitrogen gas is entrapped in the well during measurement. When preparing the solid mantle and the realization of the triple point, we fill the vacuum jacket with helium. The helium

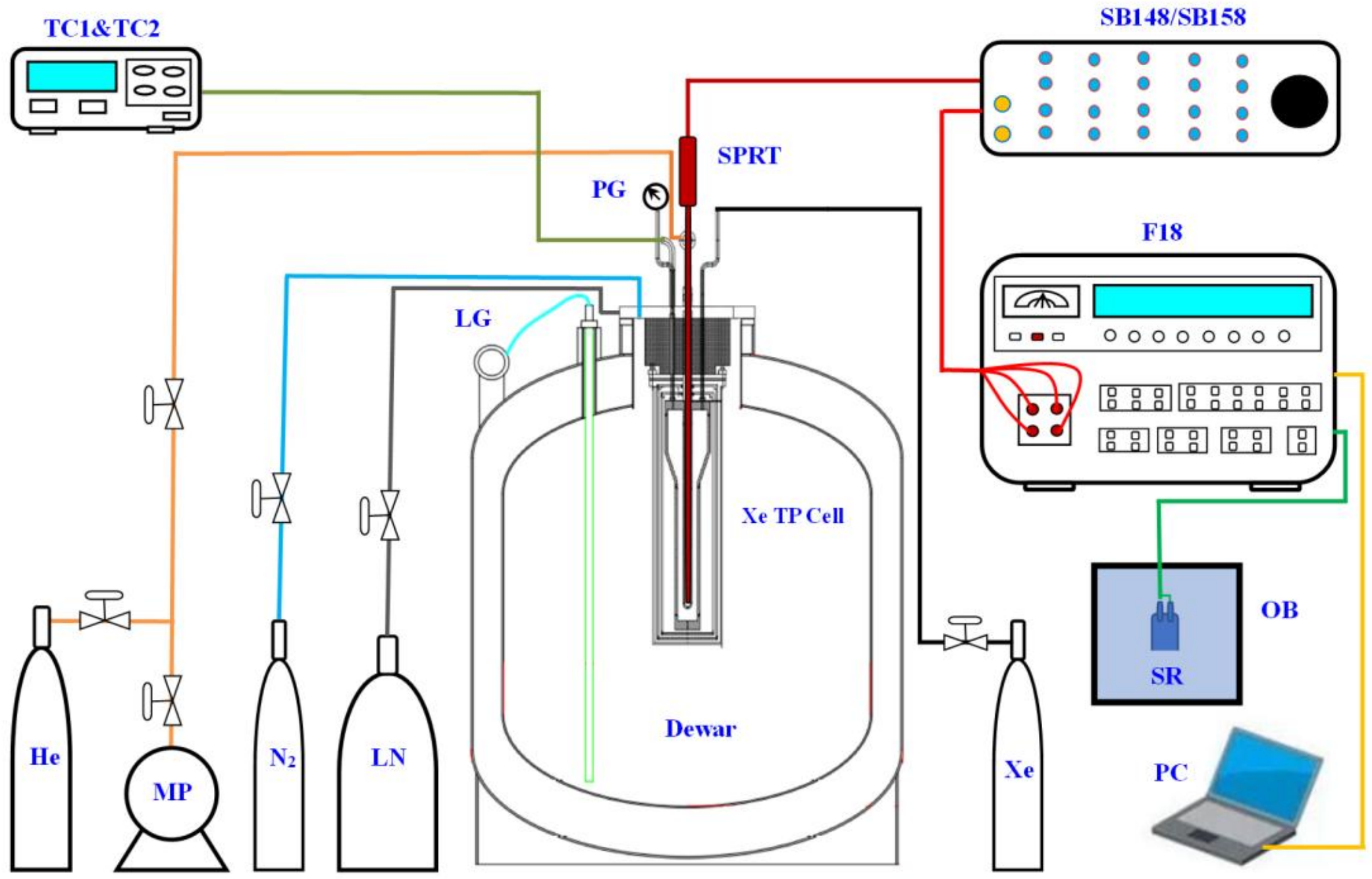

**Fig. 4.** Schematic diagram of the immersion-type Xe triple point experimental system. Temperature controller, TC1 & TC2; vacuum pump, MP; liquid nitrogen tank, LN; liquid level gauge, LG; pressure gauge, PG; ASL SB148/SB158 multi-channel switch-box, SB148/SB158; resistance ratio bridge, F18; xenon sample, Xe; standard resistor, SR; oil bath, OB.

enhances heat transfers among the cell, the radiation screens, the jacket and the liquid nitrogen.

We apply a high precision AC resistance ratio bridge, ASL F18, measuring the resistance ratios of the SPRT to the reference standard resistor (SR). The SR is the model of Tinsley 5685 A of a nominal resistance of 100 Ω. The SR has been calibrated at 20.00 °C by the primary standard of the National Institute of Metrology (NIM), China. We tabulate the calibration data in Table 3. The SR is maintained in an oil bath at 20.000 °C ± 0.002 °C. The temperature coefficient combined with the temperature uncertainty causes an error in resistance of $4.0\times10^{-9}$ Ω.

**Table 3**

Calibration of the standard resistor.

| $I$ (mA) | $t$ (°C) | $R$ (Ω) | Temperature coefficient （Ω/°C） | $U$ (Ω) /($k$ = 1) |
|---|---|---|---|---|
| 1.5 | 20.00 | 99.999 905 | $2.0\times 10^{-6}$ | $1.7\times 10^{-5}$ |

We alternatively apply two LSPRTs, Fluke no. 1733 and 1731, measuring the melting plateaus at the Xe TP in the newly developed setup. Before measurement, the thermometers were calibrated on the ITS-90 in the subrange of 83.8058-273.16 K by the primary standard of the NIM. According to the guide realizing ITS-90 [1], the resistance ratio $W=R(T_{\mathrm{fp}})/R(T_{\mathrm{tpw}})$ is measured through measuring the SPRT's resistances at the fixed points (we denote by the subscript fp), here the Ar TP and Hg TP, and the triple point of water (we denote by the subscript tpw). The ITS-90 has define the reference resistance ratios $W_{\mathrm{r}}$ at the Ar and Hg TP. Thus, the deviation $\Delta W$

$$\Delta W = W - W_{\mathrm{r}} \tag{1}$$

is obtained at each fixed point. Thus, the calibration coefficients $a$ and $b$ can be calculated out according to the interpolation equation given by the ITS-90,

$$W - W_{\mathrm{r}} = a(W-1) + b(W-1)\ln(W) \tag{2}$$

When doing calibration, we extrapolate the SPRT's resistances at the zero current by following the guide [35]. We list the data in Table 4.

**Table 4**

Calibration at the zero current for the SPRTs.

| SPRTs | $R_{\mathrm{TPW}}$ (Ω) | $W_{\mathrm{Hg}}$ | $W_{\mathrm{Ar}}$ | $a_4$ | $b_4$ |
|---|---|---|---|---|---|
| 1733 | 25.63165 | 0.84416870 | 0.21598791 | -0.00017156 | -0.00000529 |
| 1731 | 25.49349 | 0.84417357 | 0.21600809 | -0.00020372 | -0.00000946 |
| $U$(mK), $k$=2 | 0.16 | 0.59 | 0.59 | — | — |

## 3. Results

### *3.1. Experimental Procedures*

We implement a preprogram before realization of the triple point. Firstly, we pump out air from the vacuum jacket, and fill it with 4N He gas at an absolute pressure of 60 kPa. We filled in the thermometer well with $N_2$ gas at a pressure of approximately 23 kPa above the atmospheric pressure. Then, we fill in the Dewar with liquid nitrogen to the level of 90%. The jacket assembly is immersed in liquid nitrogen. We set the temperature of the controlled radiation screen at 1 K above the Xe TP for a duration at least 12 hours to ensure the Xe sample in the cell to be completely melted. Second, we reset the temperature of the screen to 0.50 K above the Xe TP. The state is maintained for a duration at least 24 hours to stabilize the Xe sample in cell. Third, we set the temperature of the screen at the supercooling temperature of 6.3 K below the Xe TP. We monitor the thermometers attaching to the outer wall of cell. Once the thermometers indicate a rapid temperature rise, we change setting the temperature to 1 K below the Xe TP until complete solidification. Fourth, we raise the temperature of the screen at 0.5 K below the Xe TP. The state is maintained for 24 hours for a relaxation of the crystal strain formed during solidification.

After finishing the preprogram, we set the temperature of the screen at 0.5 K above the Xe TP. We check the readings of the thermometers attached to the outer wall of cell. With time, the thermometers will record the temperature rise until the appearance of the melting plateau along the outer surface of the solid Xe mantle. Because of a finite heat transfer from the outer wall to the mantle center, the SPRT in the thermometer well will record the melting transition at the outer surface of mantle with 1.5 hours delay. Once the SPRT records the melting transition, we loosen the O-ring and pull out the SPRT from the well, subsequently insert a copper rod. The rod has the equal dimensions of the SPRT. We fill the well with $N_2$ gas again protecting it from icing. The rod heats the well to melt the mantle. The melt forms a liquid film between the well and the mantle, the liquid-solid interface. The consecutive liquid-solid interface facilitates our measuring the Xe TP. The initiation operation of the solid-liquid interface lasts for 10 minutes. After, we pull out the rod from the well and insert the SPRT in the well. Once again, we pump air out of the well and seal $N_2$ gas in the well at a pressure of 23 kPa above the atmospheric. We keep the far tip of the SPRT 10 mm above the bottom of the well. The operation is due to the fact that the solid phase of Xe is denser than the liquid phase. With melting, the concave dome of the solid mantle will drop away from the well tip. In contrast, the liquid-solid interface at 10 mm above the tip maintains a thin liquid film attaching to the well. Thus, the well at the position obtains a closer record of the melting transition temperature than that at the bottom tip [31].

As stated above, the Dewar is filled up with liquid nitrogen. With melting the solid mantle, the liquid nitrogen is consumed. In accordance with our multiple experiments, we find that the temperatures of the radiation screen will be unstable when the liquid level drops below 30% as stated in Section 2.4. We have tested replenishing the Dewar from below 30% to 90%. Unfortunately, the thermal inertia causes the system to lose prompt control. We find that the melting plateaus are prominently disturbed during the replenishment. In accordance to that observation, we kept without replenishing liquid nitrogen into the Dewar once finishing creation of the inner melt surrounding the thermowell in an experiment. Along this definition, we record the melting plateaus using the SPRT operating in 1 mA. We alternatively measure with the current sequence of 1 mA-$\sqrt{2}$ mA-1 mA as recommended by literature [35] to extract the measurements of SPRT at 0 mA. We report our results in 0 mA.

### 3.2. Measurement of the Xe TP

By following the experimental procedure, we conduct multiple measurements of the Xe TP. We divided the experiments in two phases. The first phase starts in the second half year of 2023 and ends at the early 2024. The second phase is in Mar. and Apr., 2025. The two phases are spaced by one year more. A single cell has been applied through the two phases. Thus, the results in the different phases facilitate a check of the long-term stability of the cell's performance. We apply the SPRT of Fluke 1731 and 1733 in measurements. We tabulate the measurements in Table 5.

**Table 5**

Measurements of the Xe TP.

| SN. | Thermometer | $\Delta T_{Avg.}$, K | $T_{tp}^1$, K ($F = 0.5$) | $T_{tp}^2$, K ($F = 0.5$) | $T_{tp}^3$, K ($F = 1.0$) | $T_{tp}^4$, K ($F = 1.0$) |
|---|---|---|---|---|---|---|
| 1 | 1733 | 0.129 | 161.40713 | 161.40573 | 161.40725 | 161.40559 |
| 2 | 1733 | 0.127 | 161.40712 | 161.40572 | 161.40752 | 161.40586 |
| 3 | 1733 | 0.136 | 161.40712 | 161.40572 | 161.40721 | 161.40555 |
| 4 | 1733 | 0.135 | 161.40707 | 161.40567 | 161.40749 | 161.40583 |
| 5 | 1733 | 0.144 | 161.40708 | 161.40568 | 161.40750 | 161.40584 |
| 6 | 1731 | 0.169 | 161.40711 | 161.40571 | 161.40737 | 161.40571 |
| 7 | 1731 | 0.148 | 161.40693 | 161.40553 | 161.40727 | 161.40561 |
| 8 | 1733 | 0.129 | 161.40706 | 161.40566 | 161.40734 | 161.40568 |
| 9 | 1733 | 0.124 | 161.40703 | 161.40563 | 161.40740 | 161.40574 |
| Mean | — | — | — | 161.40567 | — | 161.40571 |

Note: the measurements of SN.1–7 correspond to the first phase, and those of SN.8–9 correspond to the second phase; $\Delta T_{Avg.}$ labels the temperature difference between the cell's outer wall and the triple point; $T_{tp}^1$ and $T_{tp}^3$ denote the temperature measures at the Xe TP without correction of the axial heat leak; $T_{tp}^2$ and $T_{tp}^4$ denote the temperature measures at the Xe TP with correction of the axial heat leak.

For melted fraction $F$=0.5, we present the results in two records. We denote by $T_{tp}^1$ the temperature measures at the Xe TP without correction of the axial heat leak, and $T_{tp}^2$ the temperature measures at the Xe TP with correction of the axial heat leak. Similarly, for melted fraction $F$=1.0, we denote by $T_{tp}^3$ and $T_{tp}^4$ the temperature measures without and with correction of the axial heat leakage, respectively. We will present our investigation on the effect of the axial heat leak in the next section. Fig. 5 presents a typical melting plateau obtained by the immersion-type Xe TP apparatus. The plateau is measured using the SPRT 1733 on the offset temperature $\Delta T_{Avg.}$=0.144 K of the outer wall of cell. The effective plateau lasts about ten hours with a drop in liquid nitrogen from 74% to 53%. We plot the melting plateaus of all the measurements for the melted fraction $F$ dependence in Fig. 6.

As we stated above, we re-insert the SPRT after creation of the inner melt by the copper rod. The SPRT is initially warmer than the cell. We observe the SPRT's starting to drop asymptotically to equilibrium with the cell. The process generally lasts one hour for all our measurements. Thus, we choose the start point of melting plateau at the time that the SPRT gets its equilibrium with the liquid-solid interface. After, we observe the consecutively upward melting plateaus with time. The melting goes smoothly for a period. We find temperatures starting to rise rapidly with the level of liquid nitrogen approaching 50%. We reason some possibilities underlying the phenomenon. There are possible cracks forming in mantle. Cracks lead hot liquids Xe from the outer surface of mantle into the inner liquid-solid interface. The other possible reason lies in that the widening of liquid film along the well weakens the resistance against to the axial heat leak. According to general experiences, the melted fraction falls in a range from 0.75 to 0.90 corresponding to the appearance of rapidly rising rate. In this work, we account for the melted fraction 0.75 at the occurrence of rapid temperature rising. We will evaluate such an estimation effecting on the total uncertainty in Section 4. Thus, we define the point with rapid temperature rise as the end point of melting plateau. Thus, we define the effective melting plateaus between the start and the end points. As stated in Section 3.1, the inner melt is created via a copper rod. Upon the xenon mass of 1141.6 g in the cell, the transition from the solid and liquid phase needs to absorb a heat of 20130 J. The copper rod for creation of inner melt has a diameter of 8 mm. The insertion depth is 220 mm. Such a section will release a heat of 4930 J when cooled from 20 °C to the Xe TP. That heat causes a melted fraction of 0.25 of the solid xenon sample. Assume a steady heat flux. The cumulate melted-fraction at the end of the effective melting plateau is estimated of 0.72, which coincides with the experience values. We observe the temperature flatness range of the effective plateaus being in 0.37 mK–1.0 mK. The temperature flatness is the difference between the maximum and minimum values of the effective plateau.

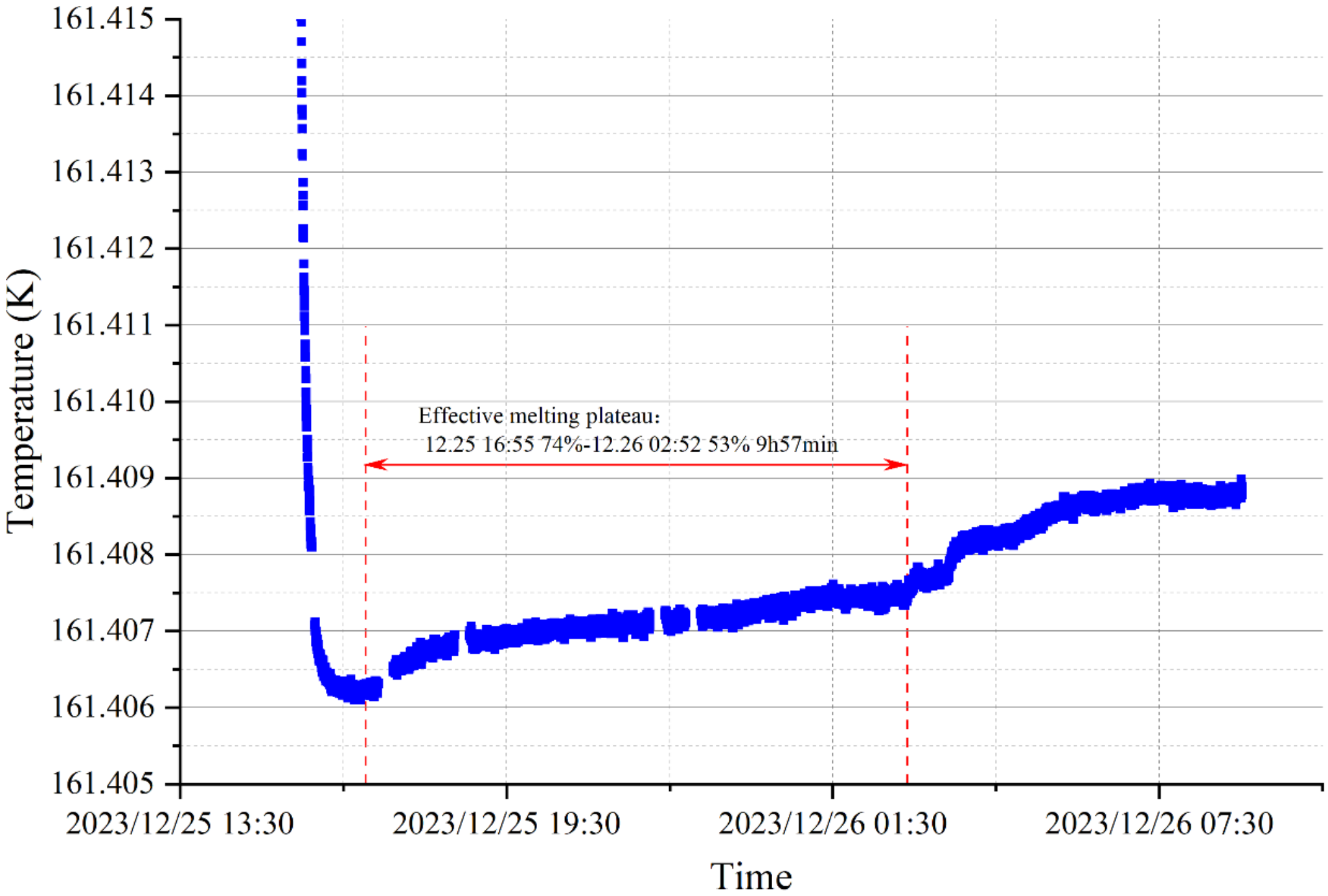

**Fig. 5.** Typical Melting plateau obtained by the immersion-type Xe TP apparatus. The liquid level was recorded of 74 % after finishing creation of the inner melt and was dropped to 53 % at the point of accelerative rise of temperature.

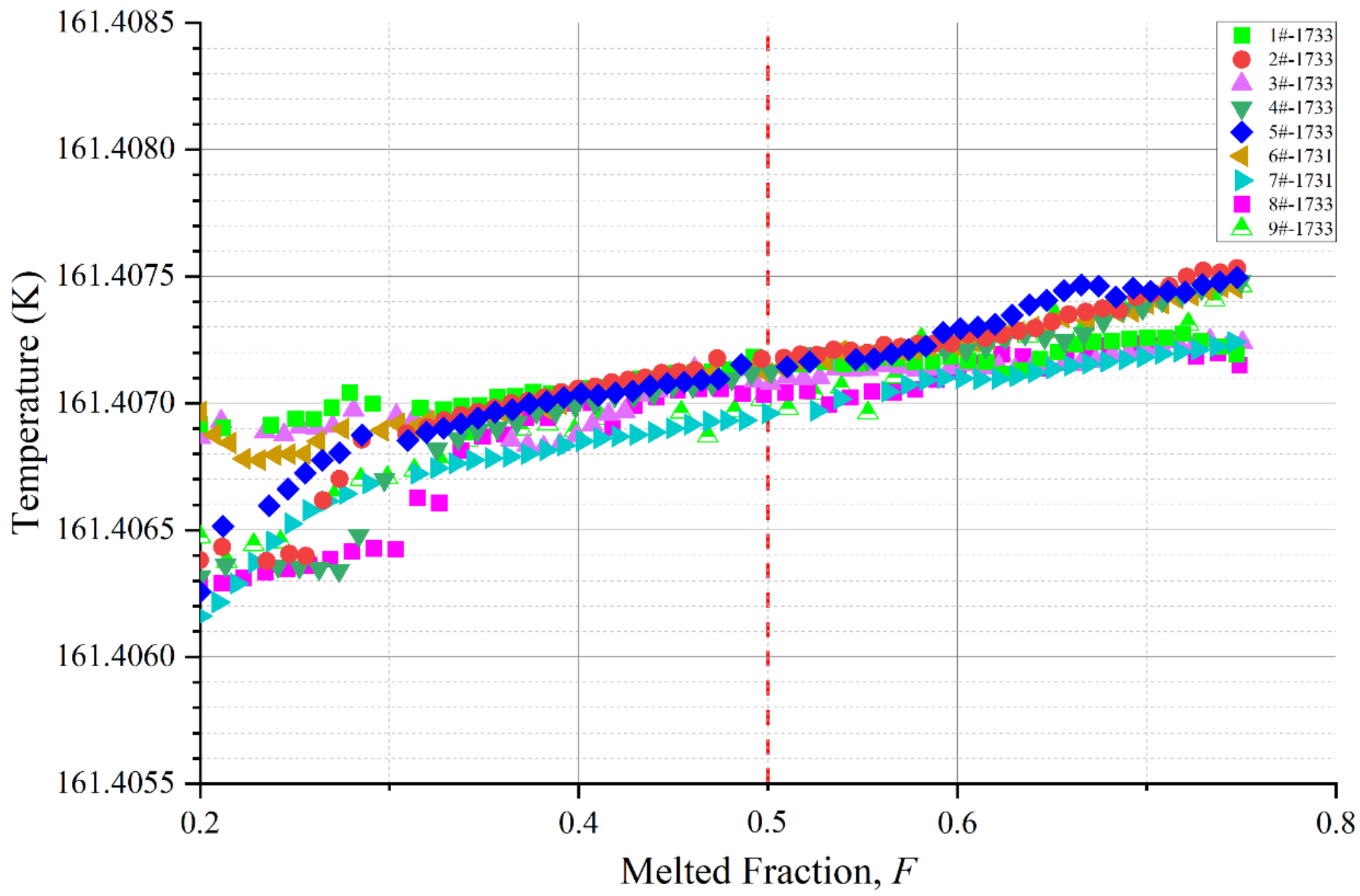

**Fig. 6.** Effective melting plateaus for the $F$ dependence at the Xe TP.

Next, we conducted a linear fit for the $1/F$ dependence of the plateaus from $F = 0.2$ to $F = 0.75$. We plotted, in Fig. 7, the melting plateaus for the $1/F$ dependence. The intercepts at $1/F = 1.0$ accounted for the values of $T_{tp}^3$ at $F$=1.0 without correction of the axial heat leak. We tabulated into Table 5 the value of $T_{tp}^3$ at $F$=1.0 with respect to each measurement.

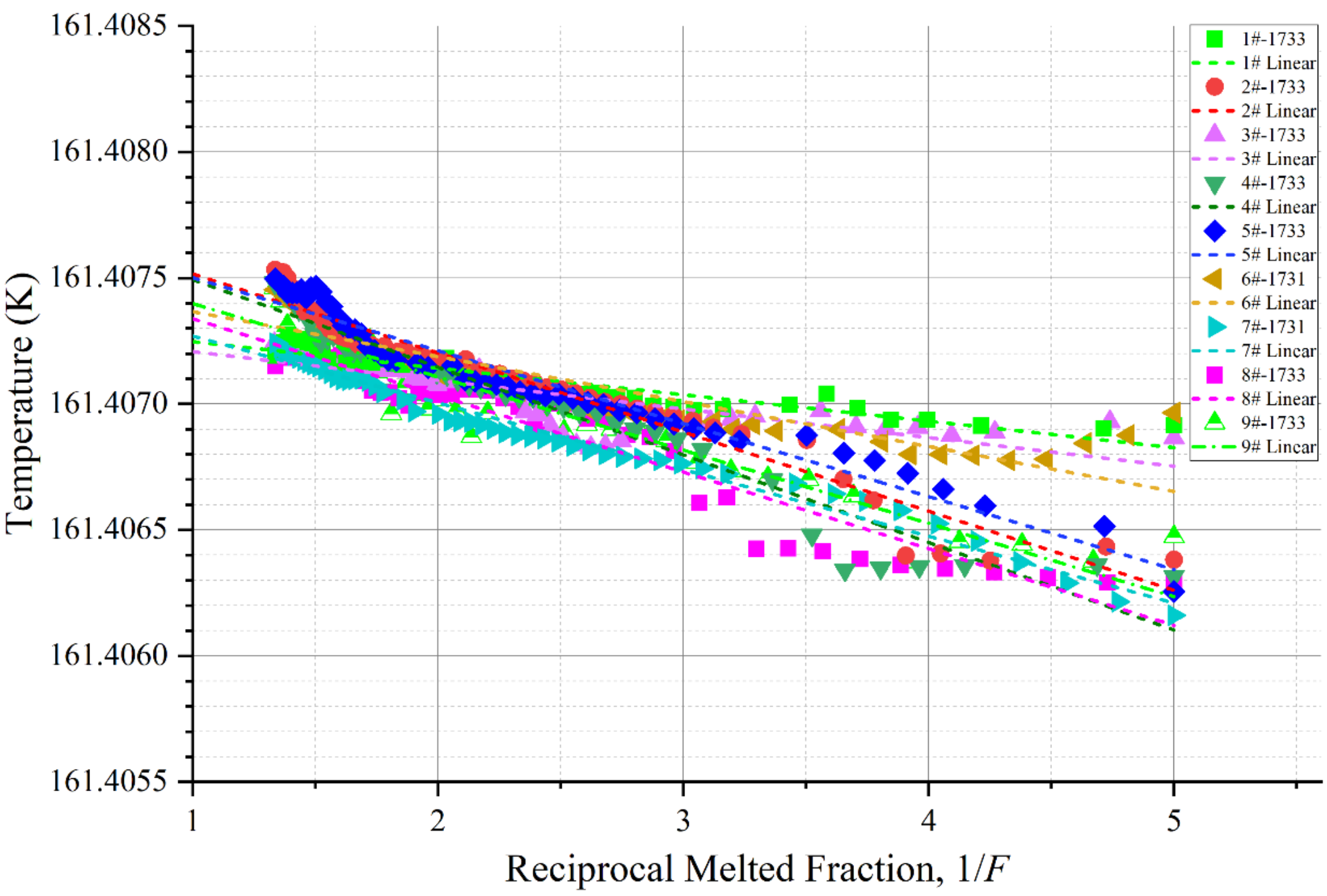

**Fig. 7.** Effective melting plateaus for the $1/F$ dependence.

### *3.3. Immersion correction*

The triple point of a pure substance is a state of the triple-phase coexistence. For the cell diagramed in Fig. 1, the triple phases are confined in a thin layer of molecular dimensions. The sensing element of a thermometer is of macro dimension. Thus, for practical measurements, the sensing element of a SPRT can only probe the liquid-solid interface instead of the triple-phase film. We stated above, the recommended depth for sensors is at least 155 mm below the triple-phase film, where the static pressure is theoretically predicted by the Clausius-Clapeyron equation,

$$\left.\frac{\mathrm{d}p_{\mathrm{mc}}}{\mathrm{d}T}\right|_{T=T_{\mathrm{tp}}}=\frac{\Delta H_{\mathrm{tp}}}{T_{\mathrm{tp}}\Delta v_{\mathrm{tp}}} \tag{3}$$

where $T_{\mathrm{tp}}$ labels the Xe TP temperature, $\triangle H_{\mathrm{tp}}$ the transition molar enthalpy between the solid and liquid phase at the Xe TP, $\triangle v_{\mathrm{tp}}$ the molar volume difference between the liquid molar volume $v_{\mathrm{l}}$ and the solid molar volume $v_{\mathrm{s}}$ [36]. The amount of $\triangle H_{\mathrm{tp}}$ is given as 17632.3 $\mathrm{J\cdot kg^{-1}}$ [19], and the value of $\triangle v_{\mathrm{tp}}$ is calculated as $5.465\times10^{-5}\ \mathrm{m^3\cdot kg^{-1}}$ [37]. The static pressure effect $\mathrm{d}p_{\mathrm{mc}}/\mathrm{d}T$ is equal to 1.999 $\mathrm{MPa\cdot K^{-1}}$. The static pressure depends on the depth in liquid. We calculate the pressure head of liquid Xe at the triple point,

$$\left.\frac{\mathrm{d}T}{\mathrm{d}l}\right|_{T=T_{\mathrm{tp}}}=\frac{-\rho g}{\left.\frac{\mathrm{d}p_{\mathrm{mc}}}{\mathrm{d}T}\right|_{T=T_{\mathrm{tp}}}}=-14.54\ \mathrm{mK\cdot m^{-1}} \tag{4}$$

where $l$ denotes the insertion depth of the sensing element of thermometer, $g$ the gravity, $\rho$ the density of liquid Xe. The pressure head is -14.54 $\mathrm{mK\cdot m^{-1}}$. The temperature value of the Xe TP is in an amount of $l\cdot(dT/dl)_{\mathrm{tp}}$ lower than the measurement of the SPRT, the insertion depth $l$ is estimated in accordance with the fraction of solid mantle. When the melting fraction $F$=0.5, the effective immersion depth of the SPRT temperature-sensing element is 220 mm. Consequently, the Xe TP temperature depressed by the pressure head is 3.199 mK.

### *3.4. Axial heat leak*

As shown in Fig. 2, two radiation screens enclose the Xe TP cell. The inner screen is passive. The outer screen is temperature controlled. When melting the solid mantle, we set the temperature of the outer screen above the Xe TP. As shown in Fig.4, the vacuum jacket has its major part immersed in liquid nitrogen. A thermally insulative layer separates the jacket from the flange of the Dewar at the room temperature. The thermometer well passes across the upper flange of jacket, the insulation layer and the flange of Dewar out to the environment in room temperature. The SPRT has its stem in the well. The temperatures between the two ends exceeds 120 K. Essential temperature gradients exist along the wall and stem. Given insufficient cooling for the wall and stem, the two components will serve as the paths conducting heat from the outer into the cell. We anticipate a tiny axial heat leaking through the paths. To quantize that effect, we check the change of the measurements of the triple point by varying the temperature of the outer radiation screen. We conduct the test by four nominal offset temperatures of 0.5 K, 0.55 K, 0.60 K and 0.70 K for the outer radiation screen.

For $F$=0.5, the corresponding average offset temperatures $\Delta T_{\mathrm{Avg}}$ at the outer wall of cell are measured to be 0.140 K, 0.153 K, 0.167 K and 0.191 K, respectively. We plot in Fig. 8 these temperature values at the triple point on dependence of $\Delta T_{Avg}$. We attempt fitting the cell measurements on dependence on $\Delta T_{Avg}$ with a linear curve and a quadratic curve regression analysis, respectively. We take into account the determination coefficient $R^2$ as the evaluator of regression model. The coefficient of determination $R^2$ is an indicator used to evaluate the goodness of fit of a regression model, and the closer it is to 1, the better the regression model fits the observed values. On the contrary, the worse the fitting degree. The linear regression analysis is shown by the dash line in Fig. 8. The fitting gives the square determination coefficient $R^2$ equal to 0.9822. The standard error of mean (SEM) of the linear regression analysis is 0.209 mK at $\Delta T_{Avg.}$= 0.140 K. A quadratic curve regression analysis, the blue line in Fig. 8, is checked. The fitting gives the square determination coefficient $R^2$ equal to 0.9907. The SEM is 0.08 mK at $\Delta T_{Avg.}$= 0.140 K. We can see that the determination coefficients of the quadratic curve regression analysis and the linear regression analysis are quite close. Its asymptotic behavior accounts for the axial heat leak vanishing with the offset temperature $\Delta T_{Avg}$. According to the test, we select the actual offset temperature $\Delta T_{Avg}$=0.140 K in our measurements of the Xe TP.

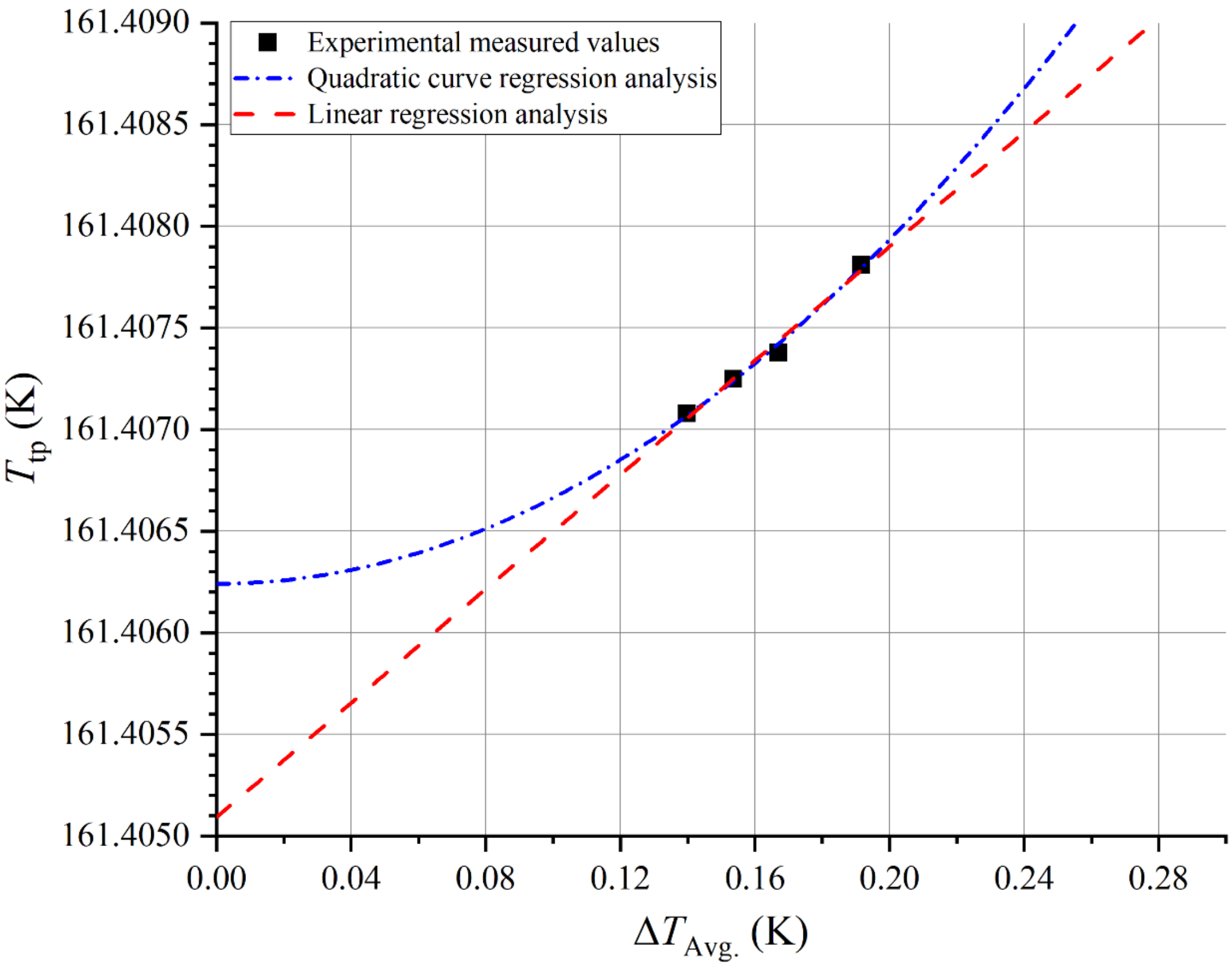

**Fig. 8.** Axial heat leak as a function of the offset temperatures $\Delta T_{Avg.}$ at $F$=0.5.

Similarly, for $F$=1.0, we plot in Fig. 9 these temperature values on dependence of $\Delta T_{Avg}$. Regarding the correction method for heat leakage, we consider that pure static heat conduction exhibits a linear relationship [38]; however, due to radial radiative heat transfer on the wall, a certain degree of nonlinearity exists. For this reason, we employed two methods to extrapolate the axial heat leak, namely quadratic curve regression analysis and linear regression analysis.

These two methods yielded two extrapolated results, which correspond to the upper limit and lower limit of the heat leak, respectively. We took into account the intermediate value as the correction factor for the heat leak effect. In this sense, the half of the difference between the upper limit and lower limit of the correction factors serves as an uncertainty measuring the Xe TP. Assume a confidence of 99 % for the uncertainty covering. The corrections using the factors cause an uncertainty of 0.186 mK and 0.227 mK with $F$=0.5 and $F$=1.0, respectively. Upon the corrections, we report the measurements of $T_{tp}^2$ and $T_{tp}^4$ by accounting for the arithmetic means over the results listed in Table 5.

In the radial direction, the solid mantle is entirely enclosed by the liquid film. In principle the temperature of the internal solid-liquid interface surrounding the thermometer well is equal to that of the external interface of the solid mantle. We may argue an entire blocking for the radial heat conduction path. Thus, we omit the effect of radial heat leak.

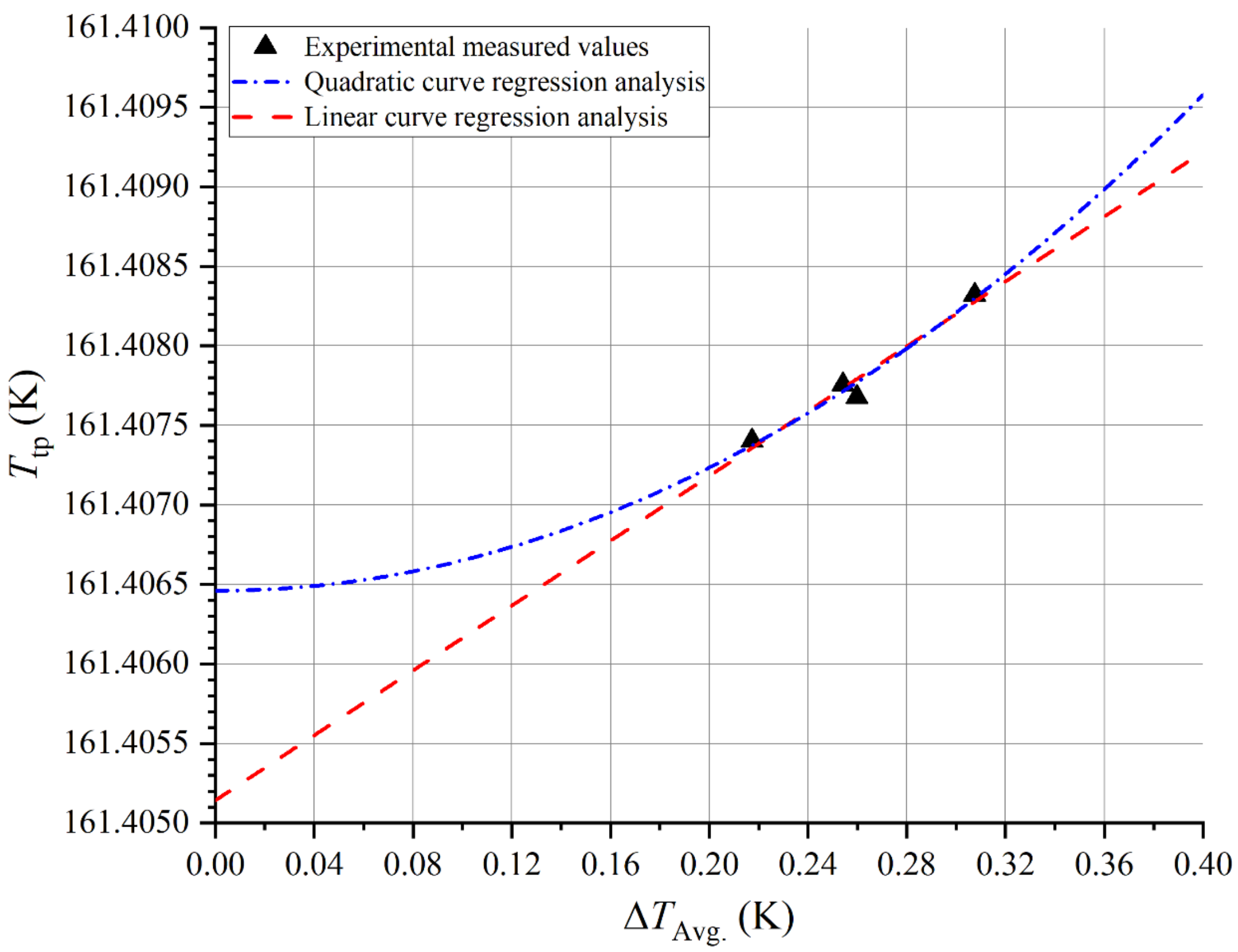

**Fig. 9.** Axial heat leak as a function of the offset temperatures $\Delta T_{Avg.}$ at $F$=1.0.

### *3.5. Repeatability and Stability*

The measurements of the Xe TP are listed in Table 5. In the table, the columns for $T_{tp}^2$ ($F$ = 0.5) and $T_{tp}^4$ ($F$ = 1.0) represent the measurements with correction of the axial heat leaks, respectively. Those results are shown in Fig. 10. All our discussion in this section focuses on $T_{tp}^2$ and $T_{tp}^4$.

We stated in Section 3.2 the measurements being divided in two phases. The first phase is accomplished using two thermometers. For the measurement result $T_{tp}^2$ at $F$=0.5, the results by the SPRT 1733 have the standard deviation of 0.026 mK. The average of 1733 is in 0.06 mK larger than the average of 1731. This agreement demonstrates a good measurement consistency of two thermometers. The second phase is accomplished using the SPRT 1733. The average of the second phase agrees in 0.06 mK lager than the average of 1733 in the first phase. That small difference accounts for a good annual stability of the cell's performance. The standard deviation for the repeatability of seven measurements using SPRT 1733 is 0.037 mK. The standard deviation of all the measurements is 0.063 mK.

For the measurement of $T_{tp}^{4}$ at $F$=1.0, the results by the SPRT 1733 have the standard deviation of 0.132 mK. We take into account this deviation for the measurement repeatability. The average of 1733 is in 0.07 mK larger than the average of 1731. This agreement demonstrates a good measurement consistency of two thermometers. The average of 1733 in the first phase is in 0.02 mK larger than the second phase. This small difference demonstrates a stable performance of the cell.

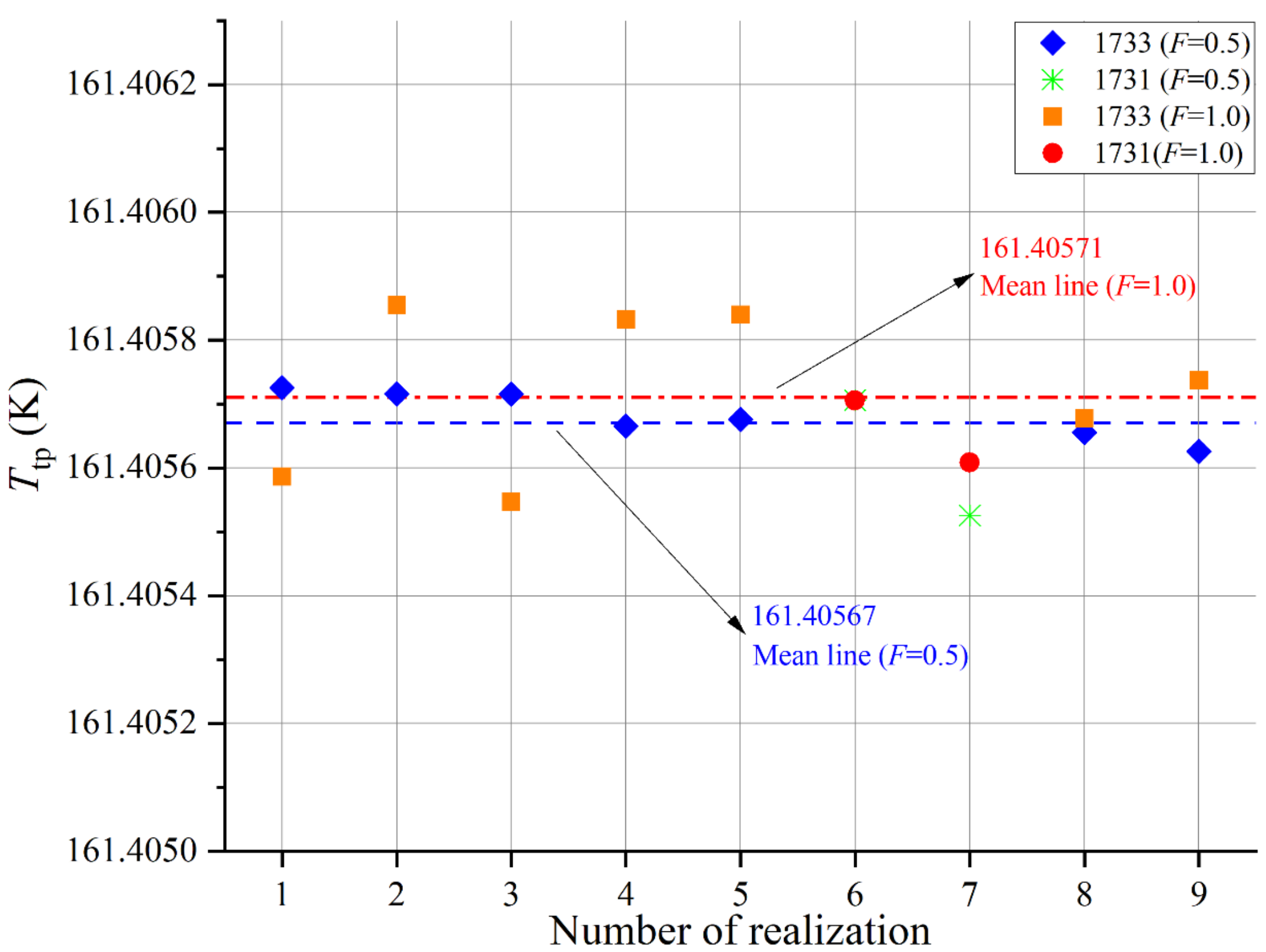

**Fig. 10.** The results of xenon triple point by correcting the thermal leakage at $F$=0.5 and $F$=1.0.

### *3.6. Determination of the Xe TP Temperature*

We reported in Table 5 the nine measurements of the temperature at the Xe TP. We extracted the temperature $T(\triangle l)$ from the fraction $F$=0.5 and $F$=1.0. The Xe TP temperature of this cell, $T_{tp}$, was calculated according to Eq. (5). For the measurement of the Xe TP, we report a new result of 161.405 67 (53) K at $F$=0.5 and 161.405 71 (55) K at $F$=1.0, respectively. The uncertainties are budgeted according to the "Guide to the Realization of the ITS-90: Platinum Resistance

Thermometry" [35], being described in the Section 4. We compare our new result with those given by the adiabatic works [19,21,23]. The comparisons are listed in Table 6. Our new result is in 0.11 mK –0.42 mK lower than those by the adiabatic works. We have no sure explaination of such difference. We argue that the uncertainty with the correction of the axial heat leak shall play an essential cause for the difference. The average values of $T_{tp}^2$ and $T_{tp}^4$ with correction of the axial heat leak in Table 5 are calculated using Eq. (5), which are the final Xe TP temperatures at $F$=0.5 and $F$=1.0, respectively.

$$T_{tp} = \frac{1}{n}\sum_{i}^{n}\left( T_i(\Delta l_i) + \frac{\mathrm{d}T}{\mathrm{d}l}\cdot\Delta l_i + \Delta T_{i,\,\mathrm{leak}} \right) \tag{5}$$

Where $T(\triangle l)$ is the temperature value at zero current calculated by the ITS-90, and $\triangle l$ is the effective immersion depth of the SPRT from the sensing element midpoint to the free surface of the liquid phase. $n$ denotes the total number of the plateaus and the subscript $i$ labels the $i$-th plateau. $dT/dl$ is the pressure-head correction coefficient. $\triangle T_{\mathrm{leak}}$ denotes the thermal leakage correction value associated with the offset of the cell surroundings from the triple-point temperature during realization.

**Table 6**

Summary of the Xe TP values reported in recent years.

| Years | Citation | Xe Purity (%) | Type of cells/PRTs | $T_{tp}$ (K) | $u(k$=1) (mK) | Notes |
|---|---|---|---|---|---|---|
| 2005 | Hill *et al.* [19] | 99.99999 | Adiabatic cell, CSPRTs | 161.405 96 | 0.32 | $F$=1.0 |
| 2014 | Steur *et al.* [21] | 99.99999 | Adiabatic cell, CSPRTs | 161.405 82 | 0.27 | $F$=1.0 |
| 2023 | Kawamura [23] | 99.9997 | Adiabatic cell, CSPRTs | 161.406 13 | 0.33 | $F$=1.0 |
| 2025 | This work | 99.9995 | Immersion cell, LSPRTs | 161.405 71 | 0.55 | $F$=1.0 |

## 4. Uncertainty evaluation

In accordance to the "Guide to the Realization of the ITS-90: Platinum Resistance Thermometry" [35], the total uncertainty for measurements of the temperature at the Xe TP can be budgeted by,

$$u_{\mathrm{total}}^2(T_{90}) = (\frac{\mathrm{d}T_{90}}{\mathrm{d}W})[u^2(W_{\mathrm{r}}) + u^2(\Delta W_{\mathrm{SRI}}) + u^2(\Delta W_{\mathrm{NU}})] \tag{6}$$

where $u_{\mathrm{total}}(T_{90})$ denotes the total uncertainty scaling in the ITS-90; $u(W_{\mathrm{r}})$ labels the uncertainty obtaining the reference resistance ratio, $u(W_{\mathrm{SRI}})$ the uncertainty by the Type 1 non-uniqueness, $u(W_{\mathrm{NU}})$ the uncertainty by the Type 3 non-uniqueness; $T_{90}$ denotes temperatures scaling in the ITS-90, $W$ the resistance ratio.

### *4.1. Uncertainty of measuring $W_{\mathrm{r}}$ at the Xe TP*

According to the guide [35], the uncertainty arising from measuring $W_{\mathrm{r}}$ at the Xe TP is given as,

$$u^2(W_r)=\frac{1}{R_{H2O}^2}[u^2(R)+W^2u^2(R_{H2O,in})+\sum_{i=2}^{N}f_i^2(W)(u^2(R_i)+W_i^2u^2(R_{H2O,in}))] \quad (7)$$

where $u(R)$ accounts for the uncertainty reading the resistance of the SPRT at the Xe TP using the apparatus; $u(R_{H2O,in})$ stands the uncertainty reading the resistance of the SPRT at the TPW with the in-situ $H_2O$ TP cell; the amount includes the contribution of the calibration uncertainty of the $H_2O$ TP cell and the standard deviation of the measurements using the SPRTs. The term $u(R_i)$ in the bracket accounts for the propagation uncertainty of calibrating the SPRT at the Ar TP and the Hg TP. The term $u(R_{H2O,\,i})$ in the bracket denotes the propagation uncertainty of calibrating the SPRT on the TPW.

In accordance with the guide, the first term in Eq. (7) constitutes of the following contributions given by Eq. (8),

$$u^2(R)=(R\frac{dW}{dT})^2u_c^2(T_{tp}(F))+R_s^2u^2(X(0))+X^2(0)u^2(R_s) \quad (8)$$

in which, the term $u_c^2(T_{tp}(F))$ is the uncertainty of measuring melting plateau; the second and third are contributions given by the electric bridge and the reference resistance. The underlying physics of Eq. (8) is the uncertainty measuring the triple point being the combination of the uncertainty by the SPRT and that by the experimental devices. The first term in Eq. (8) is,

$$u_c^2(T_{tp}(F)=u^2(T(F))+u^2(\frac{dT}{dh})\cdot h^2+(\frac{dT}{dh})\cdot u^2(h)+u^2(\Delta T) \quad (9)$$

in which, the first term on the right is the indetermination of the melting fraction; the second and the third term are the indeterminations of the correction of pressure head and the depth of liquid surrounding the thermometer well; the fourth term sums up the contributions of the repeatability (including measurement repeatability and long-term stability), the axial heat leak, the effects of impurities and isotopes. The uncertainties are listed in Table 7:

**Table 7**

Uncertainties measuring the triple point at the fraction $F$=0.5 and $F$=1.0.

| Terms | Uncertainties, mK | |
|---|---|---|
| | $F$=0.5 | $F$=1.0 |
| Indetermination for $F$ | 0.083 | 0.015 |
| Repeatability measuring $T_{tp}$ | 0.037 | 0.126 |
| Correction of pressure head | 0.000 | 0.000 |
| Immersion and heat effect | 0.067 | 0.067 |
| Axial heat leak | 0.186 | 0.227 |
| Effect of impurities | 0.055 | 0.055 |
| Effect of isotopes | 0.021 | 0.021 |
| Combination ($k$=1) | 0.225 | 0.275 |
| Expanded uncertainty ($k$=2) | 0.45 | 0.55 |

Note: the thermometer for the evaluation is 1733.

We address the information for evaluation of each term in Table 7 in the following.

First, we reason the indetermination of the melting fraction to temperature differences around the middle. We count the start point of melting plateau at the time 1.5 hours after creation of the inner melt. At the time, the plateau goes into the flat stage. According to the estimation, the corresponding melting fraction is about 0.2. When a melting plateau turns to a rapid temperature rising at the end stage of melting, the solid mantle goes incomplete. There may exist cracks in the mantle. The radial heat leak makes thee plateau depart away from the flat melting. According to the estimation, the melting fraction usually occurs between 0.75 and 0.9 at the turning point. Thus, an uncertainty arises from indetermination of the melting fraction of 0.5 at the middle point of plateau. On considering slopes of melting plateaus, we check the temperatures of 1 hour forward and backward to the middle point. We take into account the maximum temperature difference comparing to that at the middle point as the uncertainty, 0.083 mK for the plateaus being measured. Besides, when extrapolating the temperature of the Xe TP at $F$=1.0, a difference of 0.015 mK exists between the extrapolations-with melted fraction ranges of $F$=0.2-0.75 and $F$=0.2-0.9. We take into account the difference as an uncertainty to the extrapolation temperature at $F$=1.0.

Second, we have conducted seven measurements using SPRT 1733 in two phases. We take into account the standard deviation of the measures as the repeatability, which is 0.037 mK at $F$=0.5 and 0.126 mK at $F$=1.0. The repeatability is actually a combination of the short-term measurement repeatability and the long-term stability.

Third, the pressure head is a derivative from the Clausius-Clapeyron equation. We take into account a zero uncertainty for this theoretical value.

Fourth, the uncertainty for correction of pressure head arises from the indetermination of selection of immersion depth. By knowing the densities of the solid and liquid phase, the complete solid mantle has a depth of 222 mm, and the complete liquid has a depth of 238 mm in the cell. The difference is 16 mm. Given a linear melting rate, the depth at $F$=0.5 is 230 mm. We mentioned that the SPRT's tip is lift up at 10 mm. Thus, the actual depth is 220 mm. The indeterminations are restricted within ± 8 mm. On consideration of the pressure head −14.53 mKm$^{-1}$, the uncertainty doesn't exceed the range of ±0.116 mK. The range is believed of the confidence of 99%. Given canonical distribution, the respective standard uncertainty is equal to $0.116/\sqrt{3}$=0.067 mK.

Fifth, as stated in Section 3.4, the uncertainty for correction of the axial heat leak is 0.186 mK at $F$=0.5 and 0.227 mK at $F$=1.0, respectively.

Sixth, we have an analysis on the impurity and isotopic effects. We estimate the total impurity effect of 0.055 mK. As reported in the literature [25], the maximum estimated uncertainty from xenon's isotope effect is only 0.021 mK.

In accordance with the guide [35], the uncertainties contributed by the electronic device for extrapolation of the resistance ratio at the zero current are given by,

$$u^2(X(0)) = u^2(X_{\mathrm{INL}}) + \frac{I_2^4 + I_1^4}{(I_2^2 - I_1^2)^2} u^2(X_{\mathrm{DNL}}) + \frac{I_2^4}{(I_2^2 - I_1^2)^2} u^2(X_{1,\mathrm{noise}}) + \frac{I_1^4}{(I_2^2 - I_1^2)^2} u^2(X_{2,\mathrm{noise}}) + \frac{4\Delta X_{\mathrm{sh}}^2 I_2^4}{(I_2^2 - I_1^2)^2} \left(\frac{u^2(I_1)}{I_1^2} + \frac{u^2(I_2)}{I_2^2}\right) \tag{10}$$

in which, $u(X_{\mathrm{INL}})$ and $u(X_{\mathrm{DNL}})$ account for the integral and differential uncertainties of the electric bridge; they are of an equate amount of $0.25\times10^{-7}$ for measurement of resistance ratio $X$. The guide treats the equal errors generated by the current $I_1$ and $I_2$; the relative uncertainties are zero. We take into account the standard deviations of measurements of the resistance ratios by $I_1$ and $I_2$ for the uncertainties $u(X_{1,\mathrm{noise}})$ and $u(X_{2,\mathrm{noise}})$, respectively. They are $0.59\times10^{-7}$ and $0.46\times10^{-7}$. Sum up the above terms in Table 8:

**Table 8**

Uncertainties for extrapolation of the resistance ratio at the zero current.

| Terms | Uncertainties | |
|---|---|---|
| | $F$=0.5 | $F$=1.0 |
| $u(X_{\mathrm{INL}})$ | $2.500\times10^{-8}$ | $2.500\times10^{-8}$ |
| $u(X_{\mathrm{DNL}})$ | $2.500\times10^{-8}$ | $2.500\times10^{-8}$ |
| $u(X_{1,\ \mathrm{noise}})$ | $5.915\times10^{-8}$ | $5.915\times10^{-8}$ |
| $u(X_{2,\ \mathrm{noise}})$ | $4.555\times10^{-8}$ | $4.555\times10^{-8}$ |
| $u^2(X(0))$ | $1.982\times10^{-14}$ | $1.982\times10^{-14}$ |
| $R_s^2u^2(X(0))/\Omega^2$ | $1.982\times10^{-10}$ | $1.982\times10^{-10}$ |
| $u(T)$/mK ($k$=1) | 0.132 | 0.132 |

The derivative $\mathrm{d}W/\mathrm{d}T_{90}$ is 0.004152 $\mathrm{K}^{-1}$ with the SPRT 1733. The uncertainty contribution is 0.132 mK for measurement of the triple point.

The oil bath is of a temperature fluctuation of 2 mK for restoring the standard resistor. The temperature coefficient of the resistor is $2.0\times10^{-6}$ /K. Thus, the uncertainty is neglectable. The NIM's laboratory calibrating the resistor is of the identical altitude with our experimental apparatus. The oil bath of the calibration is similar with that in this work. The immersion depth in the oil bath for calibration is similar with that in this work. The effects of altitude and immersion pressure tend to zero. According to the guide, the uncertainty given by the standard resistor is:

$$u^2(R_s) = X^2(0)(u^2(R_{s,\mathrm{cal}}) + u^2(R_{s,\mathrm{ac\text{-}dc}})) \tag{11}$$

where, $u(R_{\mathrm{s,cal}})$ stands for the calibration uncertainty in DC, $u(R_{\mathrm{s,ac\text{-}dc}})$ the uncertainty because the difference between the states in AC and DC. The uncertainties are $1.7\times10^{-5}$ Ω and $0.3\times10^{-6}$ Ω, respectively. The two terms results in a contribution of 0.041 mK for measurement of the triple point.

The uncertainty for measuring the triple point of water is 0.061 mK. The total uncertainty for measurement of the triple point is budgeted in Table 9.

**Table 9**

Uncertainties for measurement of the triple point at the melting fraction $F$=0.5 and $F$=1.0.

| Term | Uncertainty, mK | |
|---|---|---|
| | $F$=0.5 | $F$=1.0 |
| Measurement of the triple point | 0.225 | 0.275 |
| Extrapolation at the zero current | 0.132 | 0.132 |
| Standard resistance | 0.041 | 0.041 |
| Triple point of water | 0.061 | 0.061 |
| Combination ($k$=1) | 0.271 | 0.314 |

### *4.2. Uncertainty Propagation*

The second term on the right side of Eq. (2) accounts for the propagation of uncertainties of the SPRT's calibrations at the Ar TP and Hg TP. The uncertainties propagate to the measurement of the Xe TP. Based on the calculations of the interpolating functions, the total propagation uncertainty is 0.394 mK. Such amount is significantly larger than the measurement uncertainty at the Xe TP.

According to the guide [35], the uncertainties of Type 1 and Type 3 non-uniqueness are calculated of 0.155 mK and 0.144 mK, respectively. Combining all the uncertainty contributions, the total uncertainty is budgeted as,

**Table 10**

Total uncertainty for measuring the Xe TP in the ITS-90.

| Term | Uncertainties, mK | |
|---|---|---|
| | $F$=0.5 | $F$=1.0 |
| Measurement of the Xe TP | 0.271 | 0.314 |
| Propagation of calibration uncertainty | 0.394 | 0.394 |
| Type 1 non-uniqueness | 0.155 | 0.155 |
| Type 3 non-uniqueness | 0.144 | 0.144 |
| Combination ($k$=1) | 0.53 | 0.55 |

## *5. Conclusion and Discussion*

We report in this article our design of a new immersion-type Xe TP apparatus for measurements of the Xe TP suitable for the long-stem SPRTs. The Xe sample used in this work is the product of Linde Corporation. The producer differs from that for the previous adiabatic works [19]. It differs literature [23] from that who supplied for Cell 1, but the same for Cell 2 [23]. We conduct realizations of the Xe TP melting plateaus in nine times using two long-stem SPRTs. The individual realization is independent. We take into account the average over the measurements as the reported result, that is 161.405 67 (53) K ($k$=1) at $F$=0.5 and 161.405 71 (55) K ($k$=1) at $F$=1.0. The experiments are divided into two phases. The SPRT 1733 are used in both phases. Thus, the results of 1733 demonstrate a long-term stability of the property of the cell. For $F$=1.0, The averages in the two phases differ by 0.02 mK. The deviation, 0.126 mK, of the seven measurements demonstrates the measurement repeatability and the long-term stability. We observe that the axial heat leaks play an essential role influencing the results.

The average without correction of the axial heat leaks is 1.66 mK above that with correction.

We compare in Table 6 our result with the previous attained from the adiabatic realizations [19][21] [23] using CSPRTs. The result in this work is lower than those of the adiabatic works in the range of 0.11 mK–0.42 mK. The studies reported in [19] and [21] have the sample of the same source. The samples claim of the purity of 7N. The samples of the two cells reported in literature [23] are supplied by two producers, the Taiyo Nippon Sanso and the Linde Corporation, respectively. The purities are reported of 5N7, which are closed to that of this work. We observe that the results reported in literature [19] and [21] agree well, being in 0.17 mK–0.31 mK smaller than that in literature [23]. Our result is 0.11 mK to 0.42 mK slightly lower than the values reported in literature [19][21] [23] using CSPRTs. We argue the possibility that impurities cause a systematic deviation between the results of the work [23] and those [19]. On considering the immersion type of our apparatus, the axial heat leaks are inevitable. Such a property is substantially larger than that in the adiabatic apparatus. We argue that the axial heat leak effect shall be an essential factor responsible for the difference between our result and those given by the adiabatic works.

During the operation of the immersion-type apparatus, we observe that the quick consumption of liquids is probably a major cause of large deviations of our measurements. The liquid nitrogen has the temperature of 84 K below the Xe TP. Thus, some significant electric power is absorbed by the radiation screen for temperature control. A major part of the electric energy is transferred to heat the liquid. The evaporation caused by the heat makes liquid drop gradually. Thus, the ambient boundary of the radiation screen is thermally instable. There are difficulties to control the screen temperatures to be stable. The melting plateaus are not much flat. The standard deviation of 0.132 mK with measurements of effective melting plateaus shall reflect the situation.

**Acknowledgments**: This work was supported in part by the National Key R & D Program of China (No.2021YFF0603804), the National Natural Science Foundation of China (Grant No. 51976206; 51861033) and the Key R & D Special Project of the of Science & Technology Department of Xinjiang Uygur Autonomous Region (No. 2020B02007).